\documentclass[pra,twocolumn,showpacs,superscriptaddress]{revtex4-1}

\usepackage{graphicx}
\usepackage{setspace}

\begin{document}

\title{Entanglement enhancement in spatially inhomogeneous many-body systems}

\author{Tobias Br\"unner}
\affiliation{Physikalisches Institut der Albert-Ludwigs Universit\"at, Hermann-Herder-Str. 3, D-79104 Freiburg, Germany}
\affiliation{Institut f\"ur Physik der Technischen Universit\"at Ilmenau, Weimarer Str. 25, D-98684 Ilmenau, Germany}

\author{Erich Runge}
\affiliation{Institut f\"ur Physik der Technischen Universit\"at Ilmenau, Weimarer Str. 25, D-98684 Ilmenau, Germany}

\author{Andreas Buchleitner}
\affiliation{Physikalisches Institut der Albert-Ludwigs Universit\"at, Hermann-Herder-Str. 3, D-79104 Freiburg, Germany}

\author{Vivian V. Fran\c{c}a}
\affiliation{Physikalisches Institut der Albert-Ludwigs Universit\"at, Hermann-Herder-Str. 3, D-79104 Freiburg, Germany}
\affiliation{Capes Foundation, Ministry of Education of Brazil, Caixa Postal 250, 70040-20 Brasilia, Brazil}
\date{\today}

\begin{abstract}
We investigate the effects of spatial inhomogeneities on the entanglement of modes of strongly correlated systems in the framework of small Fermi-Hubbard chains. We find regimes where entanglement is strongly {\it enhanced} by the presence of inhomogeneities. This contrasts recent reports of entanglement destruction due to inhomogeneities. We further study this phenomenon using concepts of Density Functional Theory and, thus, provide a general recipe for the prediction of entanglement enhancement in nanostructures. We find enhancement of up to $\sim 27\%$, as compared to impurity-free chains.  
\end{abstract} 


\newcommand{\be}{\begin{equation}}
\newcommand{\ee}{\end{equation}}
\newcommand{\bea}{\begin{eqnarray}}
\newcommand{\eea}{\end{eqnarray}}
\newcommand{\bi}{\bibitem}
\newcommand{\la}{\langle}
\newcommand{\ra}{\rangle}
\newcommand{\ua}{\uparrow}
\newcommand{\da}{\downarrow}
\renewcommand{\r}{({\bf r})}
\newcommand{\rp}{({\bf r'})}
\newcommand{\eps}{\epsilon}
\newcommand{\bfr}{{\bf r}}

\maketitle

\section{Introduction}
Nanoscale spatial inhomogeneities are ubiquitous in strongly correlated systems. Either naturally occurring, as in disordered media, or artificially prepared in nanostructures and ultracold atoms, spatial inhomogeneity can influence optical, electrical, magnetic, transport and entanglement properties \cite{sup,1,2,3,5,6,7,8,9,9a,10, pra2011,prl2008, demler, rev, imp1, imp2}. Therefore, it is of great fundamental and technological interest.

The impact of spatial inhomogeneities on entanglement, which is a key ingredient in quantum information theory, has been investigated in spin chains, Kondo and Hubbard models \cite{demler, rev, imp1, imp2, prl2008, pra2011, epl2011, coe2010}. The main goal is the development of solid-state devices for quantum information processing \cite{rev, general}, for which accessing high degrees of entanglement, despite the inhomogeneities, is essential.

For the important case of the inhomogeneous 1D fermionic Hubbard model, the {\it entanglement of modes} \cite{zanardi, dowling} or occupation-number entanglement has been studied. In this case, the ground-state entanglement between two sections (or subsystems) of the chain is quantified by the von Neumann entropy (see Eq. (2) below). The influence of several types of inhomogeneities was analyzed: localized impurities and superlattice structures \cite{prl2008}, harmonic traps \cite{prl2008, epl2011, coe2010} and disorder \cite{pra2011}. Although entanglement depends on the specific inhomogeneous component of the potential, the effect of inhomogeneities was, remarkably, found to be qualitatively the same in all cases: entanglement decreases.

All those studies focused, however, on the entanglement between a single site $-$ the smallest possible subsystem $-$ and the remaining lattice sites. The scaling relations from the single-site entanglement to the entanglement of larger subsystems, called {\it block entanglement}, have been extensively explored from the statistical point of view \cite{rev2, JPhysA2009, gu, vidal, korepin,cc, pra2008}. Nevertheless, it is still a fundamental open question whether the block entanglement is also destroyed by spatial inhomogeneities. Furthermore, from the view point of realistic applications, it is crucial to build a detailed characterization of the regimes, at which entanglement is strong regardless of unavoidable inhomogeneities.

In this paper, we address this issue by investigating the block entanglement of small 1D Fermi-Hubbard chains, where the inhomogeneity is induced by an external potential, simulating either localized impurities or superlattice structures. We find that entanglement is not always destroyed:  there are regimes where it can be even {\it enhanced} by the presence of inhomogeneities. We explore the physics behind this phenomenon from the perspective of Density Functional Theory, and estimate the optimally achievable entanglement enhancement under rather general conditions. Applying our approach to several superlattices, we predict entanglement enhancement of up to $\sim 27\%$ as compared to the homogeneous case.

The 1D inhomogeneous Hubbard model is given by \cite{hubb}
\begin{equation}
\hat{H}=-t\sum_{<ij>,\sigma}^L\hat{c}^\dagger_{i\sigma}\hat{c}_{j\sigma} + U\sum_i^L \hat{n}_{i\uparrow}\hat{n}_{i\downarrow}+ \sum_{i,\sigma}^L V_i \hat{n}_{i\sigma}\label{hm},
\end{equation}
with on-site repulsion $U$ and repulsive impurities of strength $V_i$, which we will both measure in units of the tunneling strength $t$ between adjacent sites. We adopt balanced populations ($N_\uparrow=N_\downarrow$), where $N= N_\uparrow + N_\downarrow$ is the total number of particles, $L$ the lattice size, and $n=N/L$ the filling factor. The ground-state entanglement between a block of $x$ lattice sites and the remaining $L-x$ sites is quantified by the von Neumann entropy
\begin{equation}
S_x=-\text {Tr}\left[\rho_x \log_2\rho_x\right],\label{ent}
\end{equation}
where $\rho_x=\text{Tr}_{\{L-x\}}[\rho]$ is the reduced density matrix, and $\rho$, which is calculated by exact Lanczos diagonalization, is the ground state's total density matrix. 

To set the stage, let us recall the general properties of the entanglement as defined by $S_x$: The latter depends on the degree of purity of the reduced density matrix, when it is pure ($\rho_x=|\psi_x\rangle\langle\psi_x|$), $S_x=0$; and the more mixed $\rho_x$, the greater $S_x$. That means entanglement vanishes whenever one of the possible states (in the reduced system) has unit probability, whereas the maximum entanglement is achieved when all reduced system's states are equally probable. For the infinite homogeneous Hubbard chain, in the basis of occupation number, this condition is fulfilled at half-filling ($n=1$) and $U=0$ \cite{pra2006}. Thus, for a fixed $x$ and periodic boundary conditions, conformal invariance predicts the maximum entanglement to be \cite{cc, pra2008}:
\begin{equation}
  S^{\text{max}}_{\text{hom}}(x,L\rightarrow\infty)=2+\log_2[x^{2/3}].\label{cc}
\end{equation}% For a {\it finite} chain, the highest entanglement is reached by the largest possible block, $x=L/2$ \cite{pra2008}. Thus, using the analytical result as an approximation, we estimate the optimal entanglement in finite chains as $S^{optimal}\sim S^{max}_{x=L/2}$.

%For finite chains, the entropy increases with $L$ \cite{L1,L2}, such that the upper bound for the block entanglement of a fixed $x>1$ can be estimated by $S^{max}_{x=L/2} =S^{max}_{x=1}+S^{univ}$ \cite{JPhysA2009, pra2008} in the limit of $L>>1$, where $S^{univ}$ is a universal logarithmic correction. Thus the highest possible entanglement one can reach for the block sizes investigated here are: $S^{max}_{x=2}= 2.67$, $S^{max}_{x=3}= 3.07$ and $S^{max}_{x=4}= 3.33$.

\begin{figure}[!t]
\vspace{-0.7cm}\includegraphics[width=5cm, height=3.5cm]{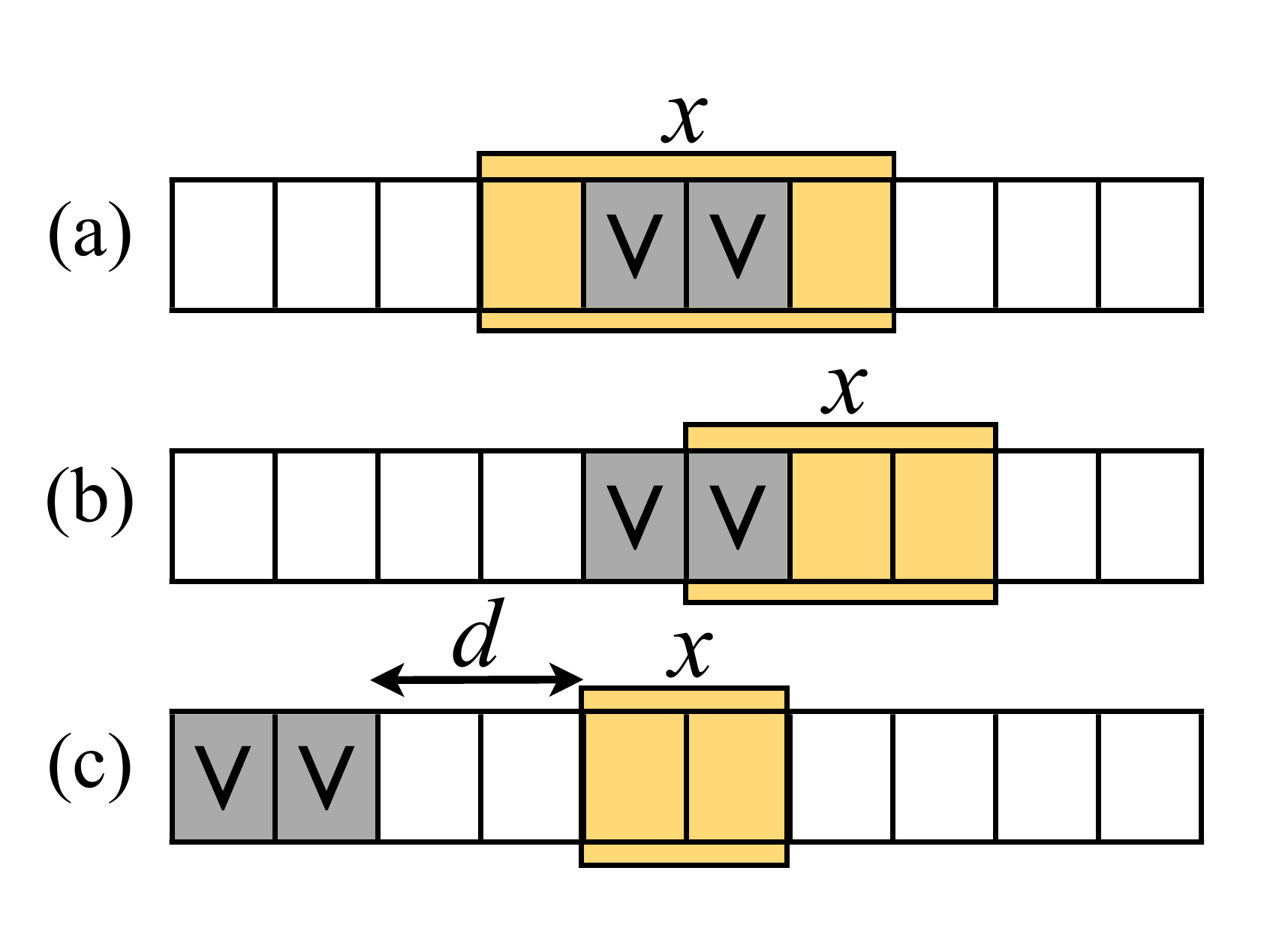}\vspace{-0.3cm}
\caption{(Color online) Illustration of the system with 2 impurities ($V$) and the relevant parameters for a block $x$: impurities (a) symmetrically distributed (2 imp sym); (b) asymmetrically  distributed (1 imp asym) and (c) at a distance $d$.}
\label{fig1}
\end{figure}

\begin{figure}[!t]
\centering
\vspace{-0.9cm}
\hspace{-1cm}\includegraphics[width=8cm, height=6.5cm]{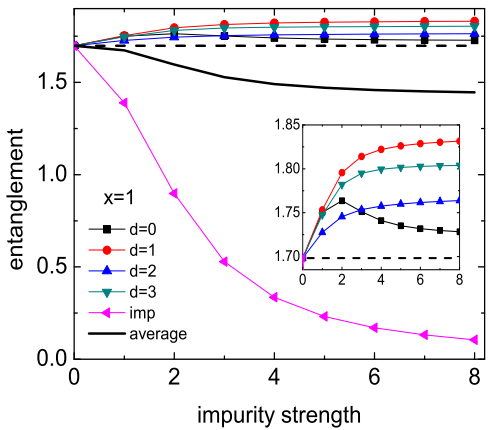}\hspace{-1cm}\vspace{-0.2cm}
\caption{(Color online) Single-site entanglement ($x=1$) as a function of the impurity strength $V/t$ for the impurity site (imp), for single sites at distance $d$ from the impurities, and the average over all possible bipartitions. Here $L=10$, $N=6$, $U=4t$, and we adopt periodic boundary conditions. Inset: enlarged view of entanglement enhancement as compared to the homogeneous case (dashed line).}
\end{figure}

\begin{figure}[!b]
\vspace{-0.4cm}
\hspace{-0.12cm}\includegraphics[width=7cm, height=5cm]{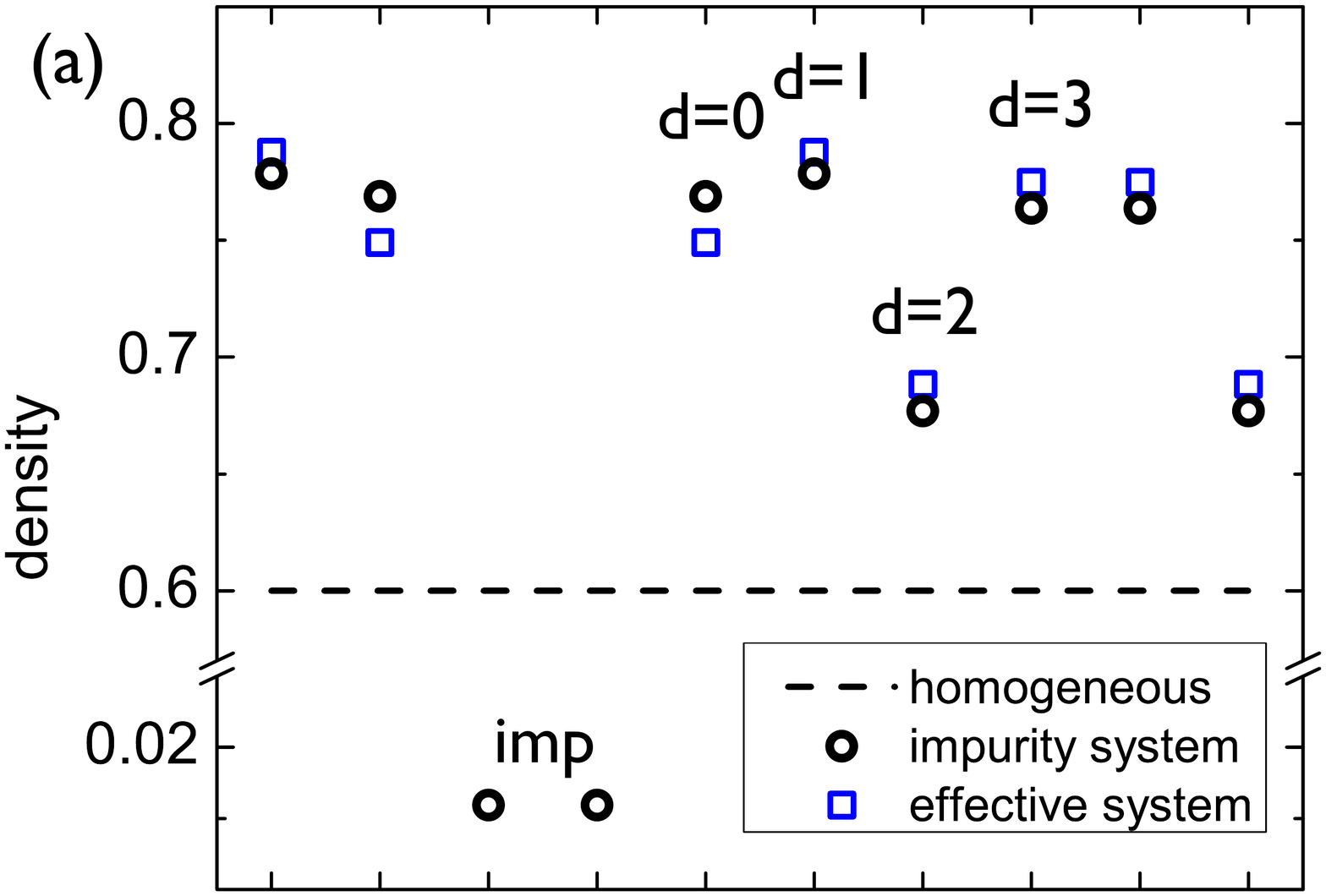}\vspace{-1.16cm}
\includegraphics[width=7cm, height=5cm]{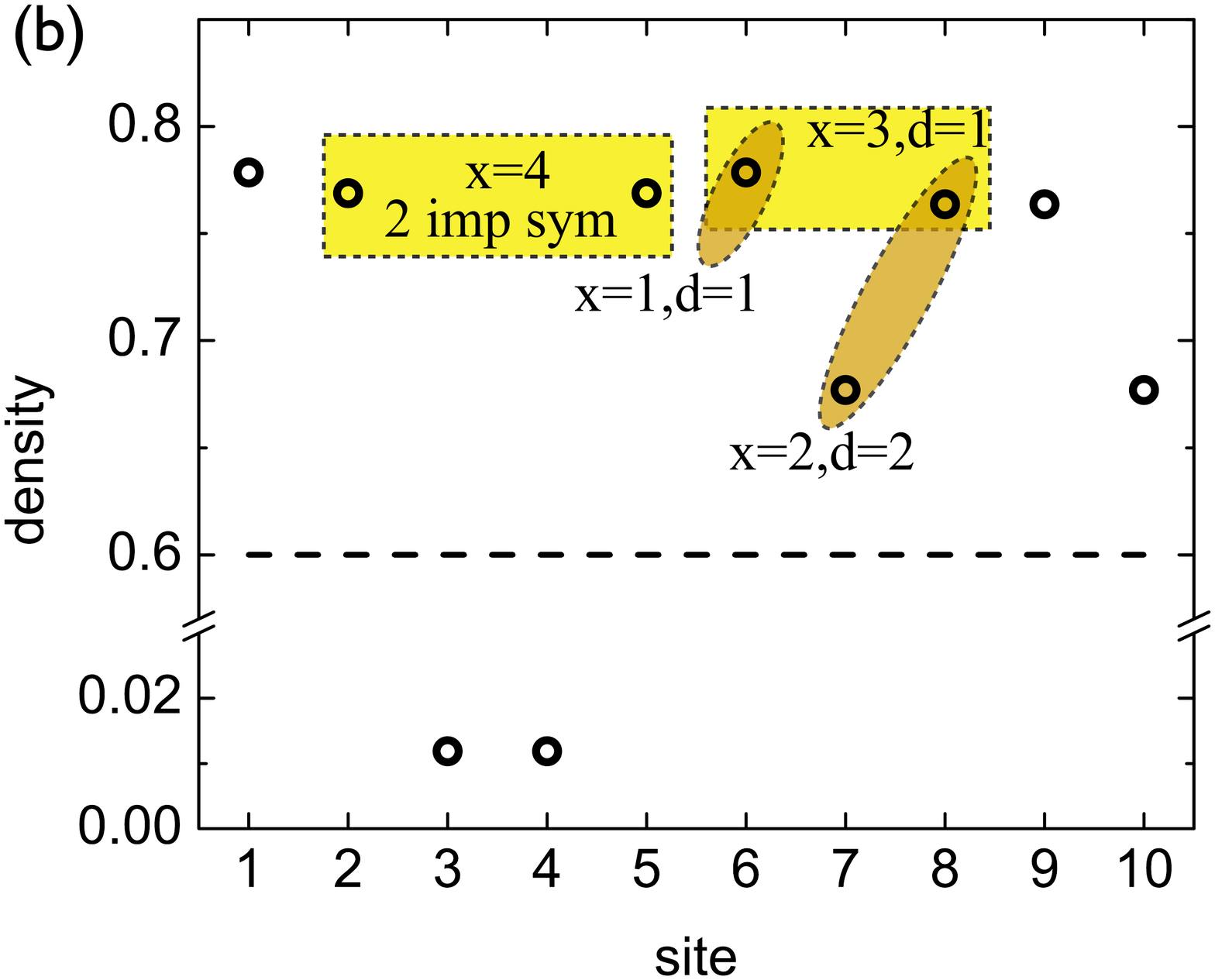}
\vspace{-0.2cm}
\caption{(Color online) Density profile of the system with $2$ impurities (see \mbox{Fig. 1}), at $V=8t$: (a) The blocks considered in \mbox{Fig. 2} are identified by $d$, and the local densities (circles) are compared to the densities in an effective chain of reduced length $L=8$, $V=0$ (squares), with otherwise identical parameters as in \mbox{Fig. 2}, and subject to open boundary conditions. (b) Predictions of the optimal entanglement enhancement for several sizes $x$ as highlighted by the coloured shapes, according to our interface-density amended LDA approach.}
\end{figure}

\begin{figure*}[!t]
\hspace{-1cm}\includegraphics[width=14.5cm]{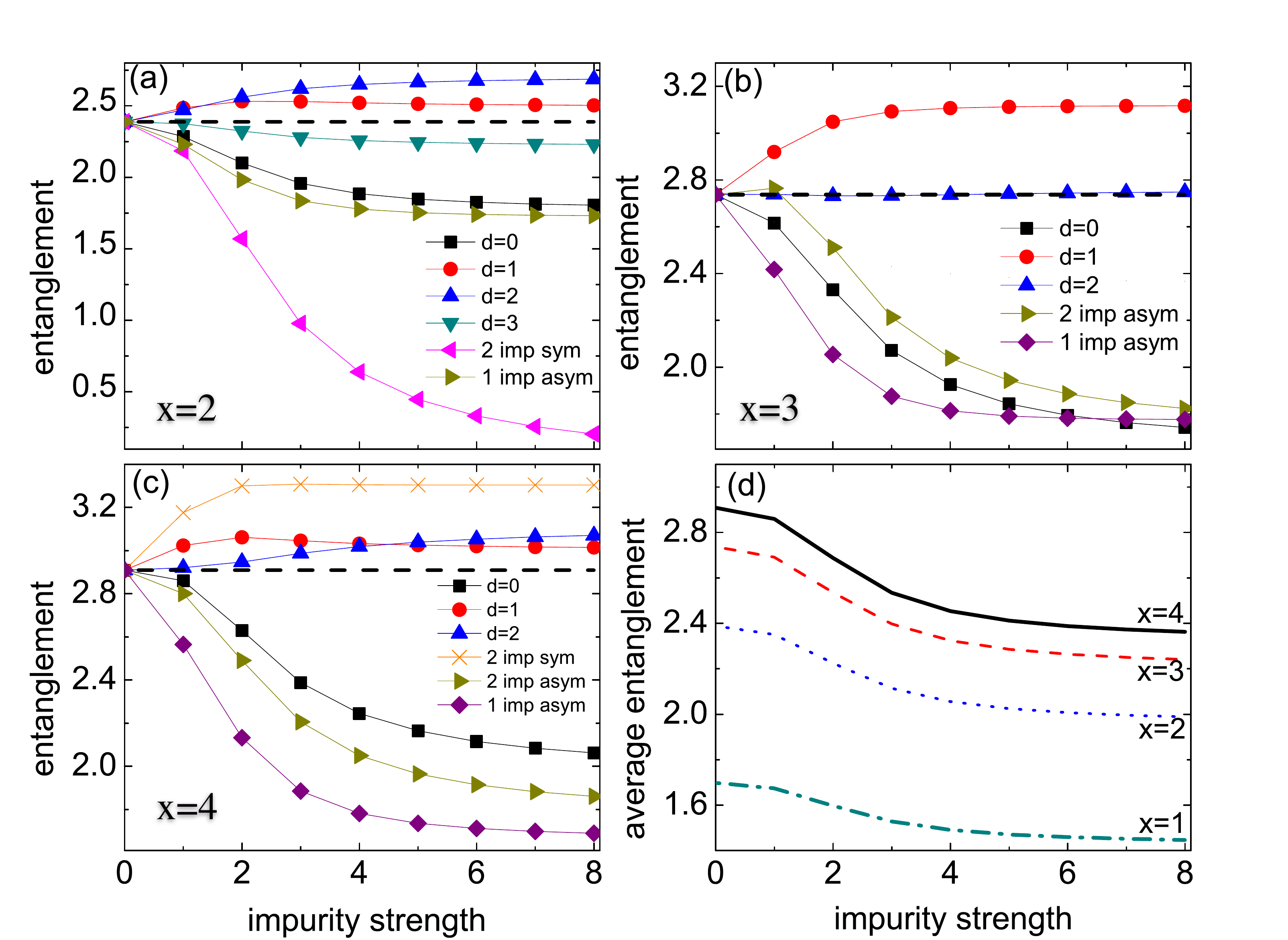}\hspace{-1.2cm}
\caption{(Color online) Block entanglement as a function of the impurity strength $V/t$ for (a) $x=2$, (b) $x=3$ and (c) $x=4$, at different distances $d$ and symmetrically (sym) or asymmetrically (asym) distributed impurities (also see \mbox{Fig. 1}). (d) Average over all possible bipartitions. $L=10$, $N=6$, $U=4t$ and periodic boundary conditions.}
\label{fig3}
\end{figure*}

\begin{figure}[ht]
\vspace{-0.2cm}
\hspace{-0cm}\hspace{0.1cm}\includegraphics[width=8.2cm]{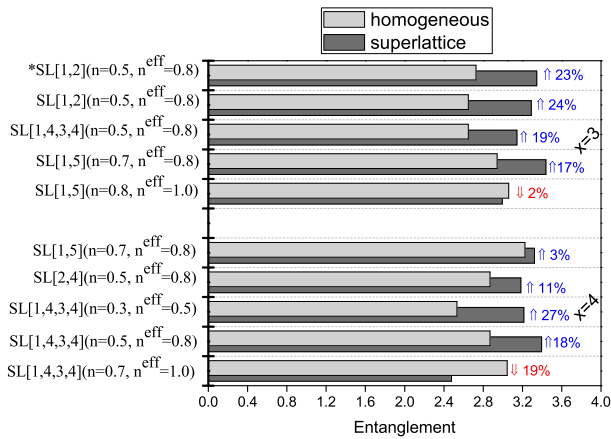}\vspace{-0.2cm}
\caption{(Color online) Entanglement for several superlattice structures for $x=3$ and $x=4$: $\text {SL}[a,\alpha,b,\beta]$ with period $[a,\alpha,b,\beta]$, where $a,b$ are sequences of sites (layers) with $V_i=V$, while $\alpha,\beta$ are layers with $V_i=0$ (for instance, the unit cell for $\text{SL}[1,1,2,2]$ and $\text{SL}[1,2]$ are $[V0VV00]$ and $[V00]$, respectively). In all cases the lattice parameters are $U=4t$, $V=8t$ and periodic boundary conditions. The entanglement enhancement is relative to the homogeneous chain ($V=0$ for all sites) of same length $L$: The upper data set ($^*\text{SL}[1,2]$) is obtained from DMRG calculations for $L=36$, while all others from Lanczos diagonalization for $L=12$.}
\label{fig4}
\end{figure}

\begin{figure}[ht]\vspace{0cm}
\includegraphics[width=8.5cm, height=5.7cm]{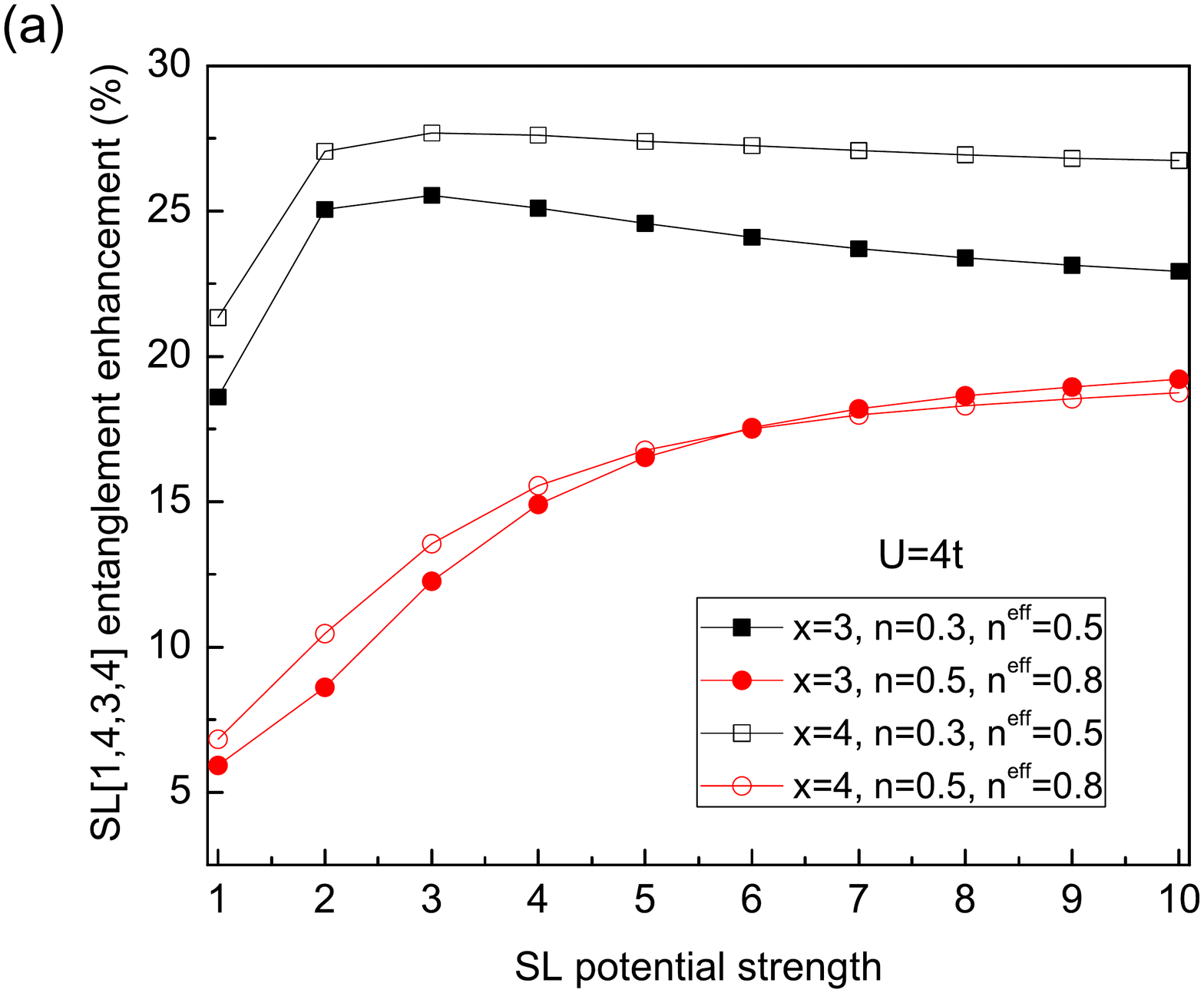}\vspace{-0.7cm}
\includegraphics[width=8.5cm, height=5.7cm]{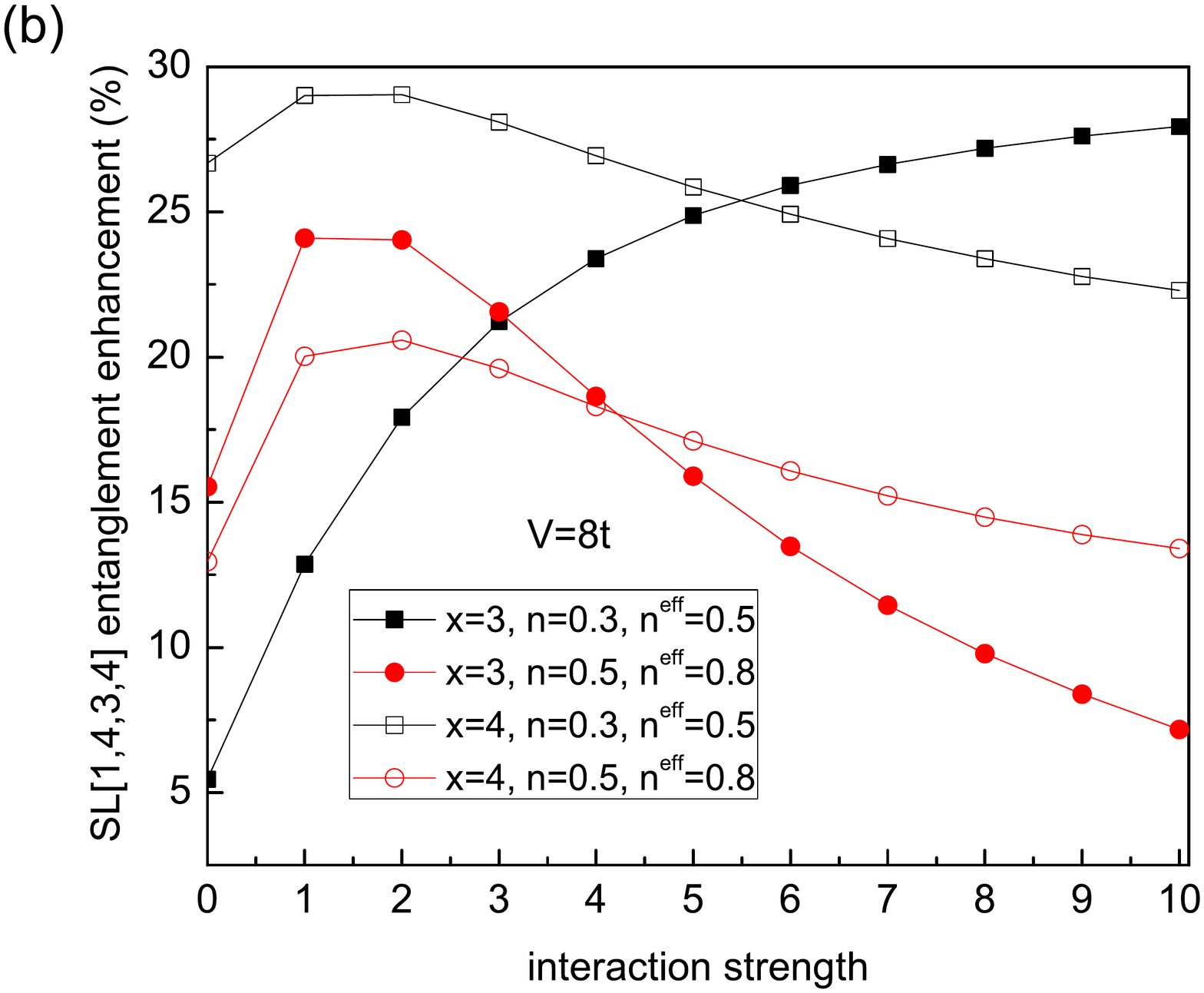}\vspace{-0.4cm}
\caption{(Color online) Entanglement enhancement in the superlattice structure $\text {SL}[1,4,3,4]$ (with unit cell [V0000VVV0000] ) for block sizes $x=3$ and $x=4$ optimally located (as proposed in Fig.3): (a) as a function of the SL potential $V/t$, for a fixed interaction $U=4t$; and (b) as a function of the interaction strength $U/t$, for a fixed potential $V=8t$. In all cases the lattice size is $L=12$, with periodic boundary conditions, and the entanglement enhancement is relative to the homogeneous chain ($V=0$ for all sites).}
\label{fig6}
\end{figure} 

\section{Localized impurities} We start by analyzing a system with $2$ neighboring repulsive impurities of same strength $V$, as \mbox{Figure 1} illustrates. We use periodic boundary conditions, such that the location of the impurities in the chain is not important. The relevant parameters are the impurities' location within the block $x$, if $x$ contains impurities; or the distance $d$ from the impurities, when $x$ does not contain impurities. 

The block entanglement is addressed from three distinct perspectives. We consider the entanglement {\it i)} between impurities and the remaining sites, the so-called impurity entanglement \cite{imp1,imp2}, and {\it ii)} between bipartitions containing both impurity and non-impurity sites. We also study {\it iii)} the average over all possible bipartitions, which quantifies the overall impact of inhomogeneities (incorporating {\it i} and {\it ii}), particularly important for the treatment of disorder. 

\mbox{Figure 2} shows the results for the single-site entanglement as a function of $V$. We find that only the impurity entanglement (case {\it i}) vanishes for $V>>t$, as the reduced density matrix of the impurity sites is pure in this limit ($\rho_x=\rho_{imp}=|0\rangle\langle0|$). For the average over all possible bipartitions (case {\it iii}), we find that the entanglement decreases with $V$, but saturates at a finite value, what is consistent with previous observations \cite{prl2008, pra2011, epl2011, coe2010}. Our results in \mbox{Fig. 2} reveal that the main contribution to this decrease comes from the impurity entanglement. Surprising, though, is the behavior in the situation when blocks do not coincide with impurity sites (case {\it ii}), as can be seen in more detail in the inset of \mbox{Fig. 2}. In this case, the degree of entanglement is not reduced. On the contrary, it is even {\it enhanced} by $V>0$. We study this interesting entanglement enhancement based on a Density
 Functional Theory (DFT) \cite{dft, dft2, dft3} point of view.  

Within DFT, all observables are functionals of the single-particle density $n(r)$ of the many-body interacting system. Unfortunately, the explicit form of the density functional of a given physical entity is, in general, unknown. However, for homogeneous Hubbard chains, where all the sites are equivalent ($n(r)=n$), the density functional for the single-site entanglement is exactly obtained from a Bethe-Ansatz solution \cite{pra2006}. One finds a broad regime of densities ($n \lesssim 0.8$ for $U>0$) where the entanglement monotonously grows with density. For inhomogeneous systems, which have by definition site-dependent densities ($n(r)=\{n_i\}$), it has been proven that the homogeneous functional is a good approximation for the single-site entanglement, within a local density approximation (LDA) replacing $n\rightarrow n_i$ \cite{prl2008,pra2011,kondo}. Consequently, the inhomogeneous single-site entanglement varies with density as in the homogeneous case \cite{pra2006}:
 monotonically increasing with density for $n_i\lesssim0.8$.

This is precisely the mechanism behind the entanglement enhancement observed in \mbox{Fig. 2}: the particles are repelled from the impurities, inducing a larger effective density at the remaining $L-2$ sites ($n^{\text{eff}}=N/(L-2)>n$), which thus leads to higher entanglement, by virtue of the above DFT argument \cite{foot2}. This can be verified at strong impurity strengths, where the impurities represent physical boundaries and therefore mimic an effective system with open boundary conditions. In \mbox{Figure 3a}, we compare the density profiles of such effective chains to our original model with impurity strength $V=8t$: the very good agreement confirms our interpretation.    

There are some additional features in \mbox{Fig. 2} which necessitate some refinement of our above LDA argument. To start with, the entanglement of blocks $d=0$ and $d=1$ differs by $\sim 6\%$ at $V=8t$ (inset of \mbox{Fig. 2}), while the difference between their densities is negligible (\mbox{Fig. 3a}). Furthermore, block $d=2$ has the smallest density among the non-impurity blocks (\mbox{Fig. 3a}), but it does not exhibit the weakest entanglement (\mbox{Fig. 2}). The most curious feature occurs for the block at $d=0$: its entanglement behaves as the other $d>0$ blocks only for very small values of $V$ and then decreases considerably (see inset of \mbox{Fig. 2}). All these features of the $d$-dependence of entanglement ought to be encoded, at least in principle, in the exact density functional for the entanglement, which is a highly nontrivial function of $\{n_i\}$.

The simplest possible functional beyond LDA is to assume that the block entanglement is not only determined by the block density itself, but is also affected by the density at its border sites, i.e., at the outmost sites of the complementing block $L-x$, in accord with the area law concept \cite{law}. In our case, the boundary surface is fixed, but weighted by the local particle density. Thus, incorporating the boundary density into our above relation between entanglement and local density, we conclude that the larger the interface density (up to $\sim 0.8$), the larger its contribution to block entanglement. 

This simple, interface-density amended LDA approach is already enough to recover qualitatively all features of \mbox{Fig. 2}. In particular, it explains the special behavior at $d=0$: one of the border sites contribution is suppressed ($n_4\sim 0$, see \mbox{Fig. 3a}), such that the entanglement saturates at a smaller value compared to all the other $d>0$ cases, in which both border sites (with non-vanishing densities) contribute to entanglement. 

Extending this analysis to $x > 1$, we can now infer which block exhibits the maximum entanglement enhancement for a given block size $x$. \mbox{Figure 3b} indicates the blocks with optimal entanglement enhancement for our case with $2$ impurities. \mbox{Figure 4} confirms this expectation: The blocks with optimal entanglement enhancement are those with the highest interface density. Qualitatively all the features observed at $x=1$ are seen again for $x>1$ (except that, for $x>2$, as the block is larger than the number of impurities available, the impurity-entanglement scenario is not defined), but the precise scaling relation from $x=1$ to larger $x$ remains non-trivial and unknown. The behavior of the average block entanglement (perspective {\it iii}), however, is similar for all $x$, as \mbox{Figure 4d} shows. This average quantity contains less details, consequently, simpler scaling properties than each particular bipartition. 

\section{Superlattice structures} We now apply the general features of optimal entanglement enhancement to superlattices (SL) \cite{sup}. Here, we consider SL structures as defined by a periodic modulation of $V_i$. From our previous analysis, one expects substantial entanglement enhancement in superlattices, whenever block and remainder are composed by border sites with $V=0$ and inner sites with $V>0$ (similar to the $x=4$ case, \mbox{Fig. 3b}). 

\mbox{Figure 5} shows the SL entanglement for blocks with this specific configuration in several distinct SL structures. We find entanglement enhancement for all cases in which the effective density ($n^{\text{eff}}=N/(L-I)$, $I$ the total number of impurities) is larger than $n$ (up to $n^{\text{eff}} \sim 0.8$), consistently with the localized impurities results. Remarkable though is the much higher relative enhancement: entanglement up to $27\%$ larger than in the impurity-free system is observed. 

The absolute entanglement values in SL are, in some cases, even {\it higher} than the maximum homogeneous entanglement one can achieve by properly adjusting the parameters, as predicted by Eq. (3) ($S_{\text{hom}}^{\text{max}}(x=3)=3.06$ and $S_{\text{hom}}^{\text{max}}(x=4)=3.33$). This certainly encourages the manipulation of nanostructures towards progress in quantum information devices.

By using Density Matrix Renormalization Group (DMRG) techniques \cite{DMRG} we have also considered a few SL entanglement of larger periodic chains (up to $L=36$). We find that the impact of $L$ on the entanglement enhancement is weak, as can be seen in Fig. 5, by comparing $\text{SL}[1,2]$ (for $L=12$) and $^*\text{SL}[1,2]$ (for $L=36$). 

We also analyze the enhancement as a function of $U$ and $V$, as shown in Figure 6. For $V>>U$, the enhancement is guaranteed for any $U$ and $V$ whenever the densities ($n$ and $n^{\text{eff}}$) are smaller than 0.8: the further from 0.8, the larger the entanglement enhancement.

%Also, by comparing the degree of entanglement of a same SL but with different $x$, we see that there are cases where $x=3$ has larger percentage enhancement than $x=4$ (see $\text{SL}^{1,3}_{4,4}$ with $n=0.5$) and situations where even the absolute value of entanglement is larger for $x=3$ than for $x=4$ (see $\text{SL}^{1}_{5}$ with $n=0.67$). These oscillations in the SL entanglement with $x$ reflect the alternating part \cite{imp1,imp2} of the entanglement entropy for open boundary conditions \cite{imp1,imp2}, here mimicked by the impurities.

In summary, our results demonstrate that spatial inhomogeneities can actually act in favor of entanglement and, accordingly, could be cleverly engineered in solid-state devices for optimal quantum information processes.

We thank Dominik H\"orndlein for providing us with his DMRG code. VF is indebted to CAPES (4101-09-0) and to the {\it F\"orderung evaluierter Forschungsprojekte} of the University Freiburg.


\begin{thebibliography}{99}
\bibitem{foot2} For $V<0$ there is no enhancement as the opposite occurs: particles (up to 4) accumulate into the impurity sites, inducing a smaller effective density $n^{\text{eff}}=(N-4)/(L-2)<n$.
\end{thebibliography}
\end{document}